\documentstyle[12pt]{article}
\newcommand{\bb}{\begin{eqnarray}}
\newcommand{\ee}{\end{eqnarray}}
\begin{document}
\rightline{hep-th/9609006}
\smallskip
\begin{center}
{\Large{\bf Understanding the Area Proposal for
Extremal Black Hole Entropy}}\\ \smallskip

A. Ghosh \footnote{e-mail amit@tnp.saha.ernet.in}
and P. Mitra\footnote{e-mail mitra@tnp.saha.ernet.in}\\
Saha Institute of Nuclear Physics\\
Block AF, Bidhannagar\\
Calcutta 700 064, INDIA
\end{center}
\smallskip\begin{abstract}
Whereas the usual understanding is that the entropy of only a
non-extremal black hole is given by the area of the horizon, there are
derivations of an area law for extremal black holes in some model
calculations. It is explained here how such results can arise
in an approach where one sums over topologies and imposes the
extremality condition {\it after} quantization.
\end{abstract}
\smallskip
It has been known for quite some time now that a black hole can be assigned
a temperature, which is a quantum effect, and is proportional to Planck's
constant. Correspondingly, there is also an entropy \cite{Bek,Hawk},
with inverse dependence on Planck's constant  and proportional to the
area of the horizon.  This entropy can be understood in a euclidean
functional integral approach \cite{GH} where the integral is evaluated
in the semiclassical approximation, {\it i.e.,} replaced by the
exponential of the negative of the minimum classical action, which
is essentially a quarter of the area.

All this is about what are now called non-extremal black holes. One
is now more often interested in a different class of black holes --
the extremal ones. These are characterized by coinciding horizons
and have qualitatively different features. Thus, the euclidean topology of
an extremal black hole is different from that of the related
non-extremal black holes. Again, the
classical action of an extremal black hole vanishes.
This results in an entropy which vanishes
\cite{HHR} or behaves like the mass \cite{GM}, but certainly does not
behave like the area.

Recently there have been some studies of black hole entropy in a string
model which count states in
what is believed to be a microscopic description
of {\it extremal} black holes and come up with a quarter of the area.
While the possibility of explicit counting is interesting, the result
(see \cite{horo})
is intriguing in view of the earlier understanding that the area formula
applies only to non-extremal black holes. It is true that
the borderline between non-extremal and extremal cases is very thin and
if one takes the extremal limit of non-extremal black holes instead of
an extremal black hole directly, one obtains the area answer. But, as
mentioned above, the euclidean topologies are different, so one should
consider not the limit but the extreme black hole by itself; and then
the string theorists' result does not match the thermodynamical answer.
A simple way out of this mismatch would be to say that one of these
calculations is wrong, but the fact that the area turns up in both the
string calculation for the {\it extremal case} and in the thermodynamical
calculation for the {\it non-extremal case} suggests that there is a
deeper truth. Clearly, for some reason, a
non-extremal case is appearing in the garb of an extremal case.
How can this have happened? This is what we seek to understand.

Usually, when one quantizes a classical theory, one tries to preserve
the classical topology. In this spirit, one usually seeks to have
a quantum theory of extremal black holes based exclusively on
extremal topologies. This leads to an entropy that vanishes or goes
like the mass, as mentioned above. Clearly, since the string model
gives  a different answer, it does not work with the extremal topology,
even implicitly. To simulate it, we shall
try out a quantization where a sum over topologies is carried out.
Thus, in our consideration of the functional integral, classical
configurations corresponding to both topologies will be included. The
extremality condition will be imposed not on the classical configurations
but on the averages that result from the
functional integration. We shall, following \cite{GH} and \cite{york},
use a grand canonical ensemble. Here the temperature and chemical
potential are supposed to be specified as inputs, and the average mass $M$
and charge $Q$ of the black hole are outputs. So the actual definition
of extremality that we have in mind for a Reissner- Nordstr\"{o}m
black hole is $Q=M$. This may be described  as {\it extremalization after
quantization}, as opposed to the usual approach of {\it quantization after
extremalization.}

The action for the euclidean version of a Reissner - Nordstr\"{o}m
black hole on a four dimensional manifold ${\cal M}$ with a boundary is
\bb
I=-{1\over  16\pi}\int_{\cal M} d^4x\sqrt gR+{1\over 8\pi}\int_{\partial
{\cal M}} d^3x\sqrt\gamma (K-K^0)+{1\over 16\pi}\int_{\cal M}
d^4x\sqrt g F_{\mu\nu} F^{\mu\nu}.
\ee
Here  $\gamma$ is the induced metric on the boundary $\partial {\cal M}$
and $K$ the extrinsic curvature. A class of spherically symmetric
metrics \cite{york} is considered
on ${\cal M}$:
\bb
ds^2=b^2d\tau^2+\alpha^2dy^2+r^2d\Omega^2,
\ee
with  the variable $y$ ranging between 0 (the horizon) and 1 (the
boundary), and $b, \alpha, r$ functions of $y$  only.  There  are
boundary conditions as usual:
\bb
2\pi b(1)=\beta, ~r(1)=r_B, ~b(0)=0.
\ee
Here  $\beta$  is the inverse temperature and $r_B$ the radius of
the boundary which will be taken to infinity.
There is another boundary condition involving $b'(0)$:
It reflects the extremal/non-extremal  nature  of  the  black  hole  and  is
therefore   different   for  the  two conditions:
\bb
{b'(0)\over\alpha(0)}&=&1{\rm ...in~non-extremal~case},\nonumber\\
{\rm but~}&=&0 {\rm ...in~extremal~case}.
\ee

The  vector
potential is taken to be zero and the scalar potential  satisfies
the boundary conditions
\bb
A_\tau(0)=0, ~A_\tau(1)={\beta\Phi\over 2\pi i}.
\ee

The  variation  of  the  action  with this form of the metric and
these  boundary  conditions  leads  to  the  Einstein  -  Maxwell
equations. The solution of a subset of these equations,
namely the Gauss law and the Hamiltonian constraint, is given by \cite{york}
\bb
{r'\over\alpha}=[1-{2m\over r}+{q^2\over r^2}]^{1/2}, ~A'_\tau=
-{iqb\alpha\over r^2},
\ee
with the mass parameter $m$ and the  charge $q$ arbitrary, except
that $|q|\le m$.
The reason why these parameters have not been
expressed as functions of $\beta,\Phi$ is that some of the equations
have not been imposed. Instead of such imposition, the
action may be simplified and then extremized with respect to
$m,q$ \cite{york}.

The value of the action corresponding to the solution depends
on the boundary condition:
\bb
I&=& \beta(m-q\Phi) -\pi (m+\sqrt{m^2-q^2})^2
{\rm ~for~non-extremal~bc},\nonumber\\
I&=&\beta(m- q\Phi) {\rm ~for~extremal~bc}.\label{I}
\ee
The first line is taken from \cite{york},
where the non-extremal boundary condition was used
in connection with a semiclassically quantized  non-extremal  black
hole. The second line corresponds to  the extremal boundary condition
used in connection with  a  semiclassically
quantized   extremal   black   hole
\cite{GMcom}.  As the euclidean
topologies  of  non-extremal  and  extremal  black   holes   are
different, quantization was done separately for the two cases in
\cite{york,GMcom}. The topology was selected before quantization.

As indicated above, a different approach is to be used here.  The
two  topologies are to be summed over in the functional integral
and the extremality condition imposed afterwards.

Thus the partition function is of
the form
\bb
\sum_{\rm topologies}\int d\mu(m)\int d\mu(q) e^{-I(q,m)},
\ee
with   $I$   given    by    (\ref{I})    as    appropriate    for
non-extremal/extremal $q$.

The  semiclassical approximation involves
replacing the double integral by the
maximum value of the integrand, {\it i.e.,} by the
exponential of the negative of the minimum  $I$.
We consider the variation of $I$ as $q,~m$ vary in both topologies.
It  is  clear from (\ref{I})
that the non-extremal action is lower than the extremal one for each set of
values of $q,~m$.
Consequently, the partition function is  to  be  approximated  by
$e^{-I_{q,m}}$,  where $I_{q,m}$ is the classical
action for the {\it non-extremal} case,
{\it minimized} with respect to $q,~m$. The result,
which should be a function of $\beta,~\Phi$, can be read  off
\cite{york}. It leads to an entropy equal to a quarter of  the
area for all values of $\beta, ~\Phi$. 
The averages  $Q, ~M$, as opposed to the parameters $q,~m$, are
calculated from $\beta,~\Phi$. We are interested in $|Q|=M$, {\it
i.e.,} the extremal black hole. This is obtained for limiting values
\bb
\beta\to\infty,~~|\Phi|\to 1,~{\rm with}~\beta(1-|\Phi|)~
{\rm finite}
\ee
for the ensemble parameters and is described by the effective action
\bb
I=\pi M^2={(\beta(1-|\Phi|))^2\over 4\pi}.\ee

It is worth emphasizing that for extremal black holes,
the parameters $\beta,\Phi$ necessarily enter in the combination
$\gamma\equiv\beta(1-|\Phi|)$
because the first law of thermodynamics takes the form
\bb
dS={dM-\Phi dQ\over T}=\beta(1-|\Phi|)dM=\gamma dM.\ee
This combination does occur here as it also does in the case
with purely extremal topology \cite{GMcom}.

Thus in the limit the partition function takes the form
\bb
Z=e^{-{\gamma^2\over 4\pi}}=e^{-\pi M^2}=e^{-{A\over 4}}.\ee
This continues to correspond to an entropy
of a quarter of the area of the horizon, which
is the value of the entropy we sought to explain.

To reach this goal, we had to define extremality {\it not}
by  equating  the  classical parameters $q,m$ before quantization,
but in terms of the
averages  $Q,M$  which  are  calculated  from  the
ensemble  characteristics  $\beta, \Phi$ and which reduce to $q,m$
for the configuration with the minimum action
in the semiclassical approximation.  It  is  because  of  this
altered definition,  and  the use of the sum over  topologies, that
non-extremal configurations have entered and we
have obtained the area law for the entropy instead of the smaller
values obtained in \cite{HHR,GM}. This suggests that the string
model  result about the entropy implicitly involves
a quantization procedure where the classical euclidean topology is ignored
and the condition of extremality imposed only after quantization.

The derivation can be translated in terms of microscopic states.
The relevant number of states can be split up into a number
of states coming from the non-extremal sector and
leading to the area formula and the small number of states coming from
the extremal sector. The second contribution can be neglected
in comparison to the first, and hence the area result survives.

It may be clarified here that we do not wish to suggest that this
is {\it the correct} way of quantization. As in other areas of physics,
there are different, often inequivalent, ways of quantization, {\it
all equally acceptable}. The
older results in quantum gravity literature correspond to quantization
with fixed euclidean topology, while the recent string counting result
agrees with, but does not explicitly involve, a sum over topologies.

It may be instructive to compare this approach with that of a recent
paper \cite{zasl} where the area law was found for the entropy
of an extremal black hole. But in that work, the {\it non-extremal} topology
was tacitly chosen without justification. As pointed out in \cite{GMcom},
the area law does not appear
if the approach of \cite{york} is adapted to the {\it extremal} topology.
The main point of the present
work is to show how one can argue for the appearance of the
non-extremal topology in the extremal case. Once the non-extremal
topology appears, the emergence of the area answer is only natural.

Lastly, it  should  be
pointed   out   that  the  functional
integral is evaluated only  in  an
approximation, but the variation of the action becomes sharp, and
the approximation better, for large black holes.
There will be corrections arising from fluctuations around the
dominant  configuration,  which  have  been  neglected  as  usual
\cite{GH}. As the
leading area term in the entropy has $\hbar$ in the denominator,
these corrections will be of order $\hbar^0$, but relatively small,
unless the area itself is comparable to $\hbar$.

\end{document}